\documentclass[12pt,twoside]{article} \input{epsf.sty}

\def\mypagenumber{1}
\def\mydate{Feb 10, 1998}
\def\myend{\end{document}}

\def\Journal#1#2#3#4{{#1}{\bf #2} (#3) #4}

\def\CQG{\em Class.\ Quant.\ Grav.}
\def\NPB{{\em Nucl.\ Phys.} B}
\def\PLB{{\em Phys.\ Lett.} B}
\def\PRL{\em Phys.\ Rev.\ Lett. }

\def\PRD{{\em Phys.\ Rev.} D}
\def\AP{{\em Ann.\ Phys.\ (N.Y.)} }

\def\CMP{\em Comm.\ Math.\ Phys. }

\def\JMP{{\em J. \ Math. \ Phys. }} 
\normalsize

\newcounter{sxn}

\newcounter{axn}

\date{}

\newdimen\mybaselineskip
\newcommand{\beeq}{\begin{equation}}
\newcommand{\eneq}{\end{equation}}
\newcommand{\be}{\begin{eqnarray}}
\newcommand{\ee}{\end{eqnarray}}
\newcommand{\bpic}{\begin{picture}}
\newcommand{\epic}{\end{picture}}

\def\dd{\partial}
\def\la{\raise.16ex\hbox{$\langle$} \, }
\def\ra{\, \raise.16ex\hbox{$\rangle$} }

\def\psibar{ \psi \kern-.65em\raise.6em\hbox{$-$} }
\def\mbar{ m \kern-.78em\raise.4em\hbox{$-$}\lower.4em\hbox{} }

\def\ep{\epsilon}

\def\n@space{\nulldelimiterspace=0pt \mathsurround=0pt }
\def\huge#1{{\hbox{$\left#1\vbox to 20.5pt{}\right.\n@space$}}}

\def\myskip{\noalign{\kern 8pt}}
\def\myeqspace{\noalign{\kern 10pt}}

\def\boxit#1{$\vcenter{\hrule\hbox{\vrule\kern3pt
    \vbox{\kern3pt\hbox{#1}\kern3pt}\kern3pt\vrule}\hrule}$}
\def\bigbox#1{$\vcenter{\hrule\hbox{\vrule\kern5pt
     \vbox{\kern5pt\hbox{#1}\kern5pt}\kern5pt\vrule}\hrule}$}

\def\ignore#1{{}}

\begin{document}

\bibliographystyle{unsrt}
\footskip 1.0cm

\thispagestyle{empty}
\setcounter{page}{\mypagenumber}

             
\begin{flushright}{\mydate ~ (.)
\\  UMN-TH-1745/98  , TPI-MINN-99/9-T \\
}
\end{flushright}

\vspace{2.5cm}
\begin{center}
{\LARGE \bf {On the Relevance of Singular Solutions}\\ 
\vspace {1.0cm}
{\LARGE  in $dS_3$ and $AdS_3$ Gravity }}\\
\vspace{3cm}
{\large Bayram Tekin\footnote{e-mail:~ tekin@mnhepw.hep.umn.edu} }\\

\vspace{.5cm}
{\it School of Physics and Astronomy, University of Minnesota}\\ 
{\it  Minneapolis, MN 55455, U.S.A.}\\ 
\end{center}

\vspace*{2.5cm}


\begin{abstract}
\baselineskip=18pt
Canonical quantization of three dimensional gravity in the first order
formalism suggests that one should allow singular solutions. This paper
addresses the importance of singular solutions in the path integral
approach to quantum gravity. Using a simple ansatz for the dreibein and
the spin connection in the de-Sitter and the anti-de-Sitter spaces 
we propose that the sum over the 3D manifolds in the path integral should
be
extended to include 2D surfaces.

\end{abstract}
\vfill



 
\newpage



\normalsize
\baselineskip=22pt plus 1pt minus 1pt
\parindent=25pt

Pure Gravity in three dimensions is a well studied
subject \cite{Deser1,Carlip1}.
When formulated in terms of the dreibein and the spin connection , it is
equivalent to Chern-Simons theory at the classical
level\cite{Townsend,Witten1} . 
At the heart of this rather curious correspondence between a spacetime 
theory of gravity and a topological field theory lies the fact that there
are only constant curvature spaces in three dimensions. There are no local
degrees of freedom both in  3D gravity and Chern-Simons theory.         

At the classical level , the formulation of gravity in terms of the spin
connection and the dreibein is equivalent to the metric formulation as
long as the dreibein is invertible. The non-polynomial dependence of
the action on the metric in the metric formalism makes the  quantization
harder if not impossible.{\footnote{ This issue resembles to the issue of
quantization of Nambu-Goto action verses Polyakov action in the bosonic
string theory.}} On the other hand canonical quantization was carried out       
in the first order formalism \cite{Witten1}. Witten showed that one has to
remove the requirement that the dreibein be invertible for the
quantization program to work.  

In a previous paper \cite{Tekin} we pointed out that, in the context of
path integral quantization, two dimensional classical solutions are
relevant in the three dimensional Euclidean gravity with zero
cosmological constant. This is possible only in the first order formalism.
Although pure gravity with zero cosmological constant was shown to
be finite and
exactly soluble {\cite{Witten1,Witten2}} the importance of the singular
(2D) 
configurations  needs to be  stressed. 
 In this paper we extend our analysis to de-Sitter and
the anti-de-Sitter spaces.

In order to define quantum theory of gravity in any dimensions, especially
as a
path integral, we have to know the global properties of the spacetimes
which can not be determined from the equations of motion. The
computation of the contributions
of different topologies to the path integral is an extremely non-trivial
problem \cite{Carlip2}  since for example in four dimensions we do not
know which
manifolds are homeomorphic. In the three dimensions we certainly have a
better understanding in the classification of the
manifolds. For a review see\cite{Rey,Benedetti,Ratcliffe}. 
The triviality of the local data in 3D gravity and a better handle
on the classification of the manifolds give us a hope to study if
the path integral quantization ( and maybe  Hartle-Hawking no initial
boundary proposal)  might be realized or not. In this paper our goal is
modest and rather pedagogical. We suggest that in the
Euclidean path integral one has to include 2D configurations which are
zero action solutions to the 3D gravity. Our effort is to show 
the relevance of singular solutions in the path integral quantization.
As stated earlier, in the canonical quantization singular solutions are
important \cite{Witten1}.

The standard definition of path integral in quantum gravity is a sum over 
the Riemannian metrics $g$ on a manifold $M$. One also has to sum over
topologically distinct manifolds. Considering a three dimensional 
Euclidean cosmological theory we have
\be
Z[\lambda] = \sum_M \int {\cal{D}}g \,\mbox{exp}{-S[\lambda,g]}  
\ee
If the Einstein-Hilbert action is used in the above formula, then
the action of a certain manifold is proportional to its volume. 
It is a non-trivial task to carry out the sum even in the
saddle point approximation. We dont know yet if the quantum theory of
gravity defined as a path integral as in the above form  makes sense or
not. We would like to stress that the sum is  understood as a sum over
3D manifolds only.     

On the other hand one might be willing to define
the quantum theory  as an integral over the spin connection and the 
dreibein.      
\be
Z[\lambda] =  \sum_M \int {\cal{D}}e {\cal{D}}w
\,\mbox{exp}{-S[\lambda,e,w]}
\label{generating2}
\ee
The lesson we learn form the canonical quantization  is that in the above
formula
we need to allow singular dreibeins  which give 2D
manifolds. This suggests that we should sum over the surfaces as well as
the 3D manifolds. 
There is certainly a bit of tension between the metric  and
the first order formulations. 
I do not know how to resolve this yet. Our analysis in this letter
is in the line of the second later formula.
 
We will consider  the following action defined on a 3D manifold $M$ with
a Euclidean signature. 
\be
S_0 = {1\over{16\pi G}} \int_{\cal{M}} d^3x \epsilon^{ijk}\Bigg\{2\,
e^a\,_i \,\dd_j\,\omega^a\,_k + \epsilon^{abc}\,e^a\,_i
\,\omega^b\,_j\,\omega^c\,_k +
{\lambda\over 3}\epsilon^{abc}\,e^a\,_i\,e^b\,_j\,e^c\,_k \Bigg\} 
\label{EH}
\ee
This action is equivalent to the
Einstein-Hilbert theory with a cosmological constant $\lambda$.  
\footnote{ In principle one can also add  $i k S_1 =
{ik\over{8\pi G}} \int_{\cal{M}}d^3x \epsilon^{ijk}\Bigg\{\,
\omega^a\,_i \,\dd_j\,\omega^a\,_k + {1\over
3}\epsilon^{abc}\,\omega^a\,_i
\,\omega^b\,_j\,\omega^c\,_k +
\lambda e^a\,_i\dd_j\,e^a\,_k
+\epsilon^{abc}\,\omega^a\,_i\,e^b\,_j\,e^c\,_k
\Bigg\}$ to $S_0$\cite{Templeton}. This action does not change the
classical equations of
motion but it will be relevant in the quantum  theory.}
The indices $(a,b,c)$ denote the tangent space and
$(i,j,k)$ denote the manifold coordinates. The metrics , $\eta_{ab}$
and $g_{ij}$ have Euclidean signature. $\lambda < 0 $ corresponds to the
de-Sitter and $\lambda > 0$ to the anti-de-Sitter space. The ``dual"
Riemann tensor can be defined to be $R^a\,_{kj} = \dd_k \omega^a\,_j
-\dd_j
\omega^a\,_k +\epsilon^a\,_{bc}\omega^b\,_k \omega^c\,_j $. The relation
between the
Ricci tensor and the dual Riemann tensor is 
$R_{ij} = e^a_i E^k_b \epsilon_{abc}R^c\,_{jk}$.  $E^k_b(x)$ lives in the
cotangent bundle as usual. The equations of motion demand that $R_{ij} = -
2\lambda g_{ij}$ and so the scalar curvature is $R = -6 \lambda $. We
are interested in the singular solutions as well as regular ones.

We adopt the following  simple $SO(3)$ symmetric ansatz which can be
rather loosely called monopole-instanton.{\footnote{For an interesting 
ansatz see the paper\cite{Ramond}.}         
\be
&&e^a\,_j(\vec{x})= {G\over r} \left[ -\epsilon^a\,_{jk}\,
\hat{x}^k\,\phi_1 + \delta^a\,_j\,\phi_2 +(r A-\phi_2)\,\hat{x}^a
\hat{x}_j \right]\\          
&&w^a\,_j(\vec{x})= {1\over r} \left[ \epsilon^a\,_{jk}\,
\hat{x}^k\,(1-\psi_1)+ \delta^a\,_j\psi_2 +(r B-\psi_2)\hat{x}^a
\hat{x}_j\right]
\label{ansatz}
\ee
The functions $A$, $B$, $\phi_i$ and $\psi_i$ depend on $r$ only. What we
mean by $r$ should be clear from $ r^2= \eta_{ij}x^i x^j$. 
The symmetric ansatz is non-trivial since the spacetime is
constructed by a collaboration of the dreibein and the spin connection
which are coupled through the equations of motion.

The metric in the manifold can be recovered by the relation 
$ g_{ij} = \eta_{ab} e^a\,_i e^b\,_j$ which yields; 
\be
g_{ij} = {G^2\over r^2}\Bigg\{ (\phi_1^2 + \phi_2^2)
(\delta_{ij} -\hat x_i \hat x_j)
         + r^2 A^2 \hat x_i \hat x_j\Bigg\}   
\ee
The Riemann tensor is calculated to be
\be
R^a\,_{ij} &=&
 {1\over r^2}
  \ep_{ij b}\,  \hat x^a \hat x^b\, (\psi_1^2 + \psi_2^2 -1)
+ {1\over r} (\ep^a\,_{ij} -  \ep_{ij b}  \hat x^a \hat x^b )
(\psi_1' + B\psi_2) \cr
\noalign{\kern 10pt}
&& \hskip 4cm + (\delta^a\,_j \hat x_i - \delta^a\,_i \hat x_j)
{1\over r}(\psi_2' - B\psi_1)
\label{Riemann tensor}
\ee

The action reduces to the following form
\be
S_0 = - \int_0^R dr \, \Bigg\{
 \psi_a' \epsilon_{ab}\phi_b +B\psi_a \phi_a
+ {A\over 2}(\psi_a \psi_a +\lambda G^2 \phi_a\phi_a -1) \Bigg \}
\label{action}
\ee
where $\{a,b\}=(1,2)$ and $ \epsilon_{ab}$ is antisymmetric. Summations
are implied over the repeated indices. The upper limit $R$ is arbitrary
for now.
Varying the action with respect to six fields give the equations of
motion. 

\be
\label{eqn1}
\epsilon_{ab}\phi_b' - A\psi_a - B\phi_a  = 0 \\
\label{eqn2} 
\epsilon_{ab}\psi_b' - B \psi_a -\lambda G^2A\phi_a = 0 \\
\label{eqn3}
\psi_a \psi_a + \lambda G^2\phi_a\phi_a - 1 = 0 \\
\phi_a \psi_a = 0 \\
\ee
The general solutions of these equations can be easily found as 
\be
\psi_1 = {1\over\sqrt{1+\lambda G^2 f^2(r)}}\cos\Omega(r)&& \hskip 1 cm
\psi_2 = {1\over\sqrt{1+\lambda G^2 f^2(r)}}\sin\Omega(r) \\
\phi_1 = f(r){1\over\sqrt{1+\lambda G^2 f^2(r)}} \sin\Omega(r)&& \hskip 
0.2cm 
\phi_2 = - f(r) {1\over\sqrt{1+\lambda G^2 f^2(r)}}\cos\Omega(r)\\ 
A= - {f(r)'\over {1+\lambda G^2 f^2(r)}}&&\hskip 2.5 cm  B= \Omega'(r) 
\ee 
The two functions , $f(r)$ and $\Omega(r)$ , can not be determined from
the equations. Both of them represent the gauge
degrees of freedom in the theory.
The metric becomes
\be
g_{ij} =  G^2\Bigg\{(\delta_{ij} - \hat{x}_i\hat{x}_j)
{f^2\over {r^2(1+\lambda G^2 f^2)}} + {f'\,^2\over{(1+\lambda G^2
f^2)^2}} \hat{x}_i\hat{x}_j \Bigg\}
\label{metric}
\ee
The line element in polar coordinates is
\be
(d s)^2 = {G^2\over{1+\lambda G^2 f^2}}\Bigg\{ f^2 d\Omega_2 +
{1\over{1+\lambda G^2 f^2}} (d f)^2 \Bigg\}
\ee  
Before we move any further let us mention that choosing the
conformal coordinate $ f(r) = 4r/G(4-\lambda r^2)$ one obtains the
standard $dS_3$ and $AdS_3$ vacuum solution.
\be
g_{ij} = {16\over(4 +\lambda r^2)^2} \delta_{ij} 
\ee
In the quantum theory action is the relevant quantity. An easy computation
gives
\be
S_0 = - \lambda G^2 \int_0^R dr {f^2 f'\over (1+\lambda G^2 f^2)^2} =
-{\lambda \over {4\pi G}}\int d^3r \mbox{det e} =
-{\lambda\over{4\pi}}\mbox{Vol}M
\ee 
Volume of the $dS_3$ and $AdS_3$ are infinite. We refer the reader to the
discussion of Gibbons-Hawking \cite{Gibbons} on how to remove the 
divergence in this context. Recently in the line of AdS/CFT
correspondence  Witten \cite{Witten3} gave a description to interpret
these divergences as the counter terms of the conformal field theory
living in the boundary ($S^2$ in this case). Our goal is different. We are
interested in the  finite action solutions.
The singular gauge $f(r) = C $ , where $C$ is a constant,  has zero
action. This corresponds to a two dimensional sphere with a constant
radius depending on the choice of $C$. These solutions should be included
in the sum for the path integral Eqn (\ref{generating2}). 

We have proposed  that two dimensional solutions, having zero action, are
relevant for three
dimensional quantum gravity defined as a path integral over the dreibein
and the spin connection. Our ansatz was rather simple. This work
can be generalized to arbitrary two dimensional surfaces. 
A word is in order regarding the loop computations. In the case 
of vanishing cosmological constant Witten \cite{Witten1,Witten2} had
calculated the generating functional exactly for a given manifold $M$.    
The result reduces to the calculation of the Ray-Singer torsion of the
manifold. It is not clear to the author now how to compute this torsion
for singular solutions. For the case of non-zero cosmological constant 
an exact solution has not been given yet. 

\leftline{\bf Acknowledgements}

I would like to thank Yutaka Hosotani and Steven Carlip for useful
discussions. This work was supported by the Graduate School of University
of Minnesota through the Dissertation Fellowship and by the U.S.
Department of Energy under the contracts DE-FG02-94ER-40823.

\vskip 1cm

\leftline{\bf References}  

\renewenvironment{thebibliography}[1]
        {\begin{list}{[$\,$\arabic{enumi}$\,$]}  
        {\usecounter{enumi}\setlength{\parsep}{0pt}
         \setlength{\itemsep}{0pt}  \renewcommand{\baselinestretch}{1.2}
         \settowidth
        {\labelwidth}{#1 ~ ~}\sloppy}}{\end{list}}

\myend
\begin{thebibliography}{99}

\small

\bibitem{Deser1}
S.\ Deser ,R.\ Jackiw and G.\ 't Hooft ,\Journal {\AP}{152}{1984}{220}
; G.\ 't Hooft ,\Journal {\CMP}{117}{1988}{685}
;S.\ Deser and R.\ Jackiw  ,\Journal {\CMP}{118}{1988}{495}

\bibitem{Carlip1} 
S. \ Carlip ``Quantum Gravity in 2+1 Dimensions" Cambridge
University Press (1998)

\bibitem{Townsend}
A.\ Achuc\'{a}rro and P.K.\ Townsend, 
\Journal{\PLB}{180}{1986}{89}

\bibitem{Witten1} 
E.\ Witten, \Journal{\NPB}{311}{1988}{46}

\bibitem{Witten2}
E.\ Witten, \Journal{\NPB}{323}{1989}{113}


\bibitem{Tekin}
B.\ Tekin ``Monopole-Instanton Type solutions in 3D Gravity"  preprint
hep-th/9812083   

\bibitem{Carlip2}
S. \ Carlip ,\Journal{\PRD}{46}{1992}{4387}; \Journal{\CQG}{10}{1993}{207}
; \Journal{\CQG}{15}{1998}{2629}


\bibitem{Rey}
M. \ Lachi$\grave{e}$ze-Rey , J-P. \ Luminet ``Cosmic Topology"
preprint gr-qc/9605010 

\bibitem{Benedetti}
R. \ Benedetti , C. \ Petronio " Lectures on  Hyperbolic Geometry"
Springer-Verlag 1992


\bibitem{Ratcliffe}
J. G. \ Ratcliffe " Foundations of  Hyperbolic Manifolds" Springer-Verlag
1994


\bibitem{Templeton}
S.\ Deser , R.\ Jackiw and S.\ Templeton ,\Journal{\PRL}{48}{1982}{976}

\bibitem{Ramond}
B. \ Keszthelyi, E. \ Piard , P. \ Ramond , \Journal{\JMP}{32}{1991}{1}

\bibitem{Gibbons}
G. W. \ Gibbons , S. W. \ Hawking , \Journal{\PRD}{15}{1977}{2752}

\bibitem{Witten3}
E. \ Witten  \Journal{Adv. Theor. Math. Phys}{2}{1998}{505}

\end{thebibliography}
